\documentclass[preprintnumbers,amsmath,amssymb,prd]{revtex4}
\usepackage{graphicx}
\begin{document}
\preprint{}
\title{Decay of the Cosmological Constant. Equivalence of Quantum Tunneling and Thermal Activation
in Two Spacetime Dimensions}
\author{Andr\'es Gomberoff}
\affiliation{Centro de Estudios Cient\'{\i}ficos (CECS), Valdivia, Chile.}
\author{Marc Henneaux}
\affiliation{Physique Th\'eorique et Math\'ematique, Universit\'e
Libre de Bruxelles \& International Solvay Institutes, ULB Campus Plaine C.P. 
231, B--1050 Bruxelles,
Belgium, and \\Centro de Estudios Cient\'{\i}ficos (CECS),
Valdivia, Chile.}
\author{Claudio Teitelboim}
\affiliation{{Centro de Estudios Cient\'{\i}ficos (CECS), Valdivia, Chile.}}
\begin{abstract}
We study the decay of the cosmological constant in two spacetime dimensions
through production of pairs. We show that the same nucleation process looks as quantum
mechanical tunneling (instanton) to one Killing observer and as thermal activation
(thermalon) to another. Thus, we find another striking example of the deep 
interplay between gravity, thermodynamics and quantum mechanics which becomes 
apparent in presence of horizons.
\end{abstract}

\maketitle

\section{Introduction and Summary}

The need to reconcile the observed small value of the
cosmological constant with the value that standard elementary
particle theory predicts is a major challenge of theoretical
physics \cite{Weinbergetal}. A number of years ago, a possible
mechanism for relaxing $\Lambda$ was proposed \cite{BT}. Its
essential aspect is the nucleation of membranes (domain walls).
The cosmological constant, which becomes a dynamical field, jumps
across the membrane and it is decreased inside it. In the version
of the mechanism proposed in \cite{BT}, the membranes were
nucleated through quantum-mechanical tunneling, for which the path
integral is dominated by an instanton solution of the Euclidean
equations of motion. Recently \cite{GHTW}, a novel variant of this
mechanism has been proposed in which the membranes are created
through classical thermal effects of de Sitter space rather than
quantum-mechanically (see \cite{Garriga,Hackworth} for related recent discussions). 
The Euclidean solutions relevant to this
process were named ``thermalons". These new solutions are
time-independent in contradistinction with the instantons, which
are time-dependent.

In order  to obtain additional insight into the thermalons,
we shall examine their analog in the simple possible context of
two-dimensional spacetime.
In doing so we encounter
yet another fascinating instance of the subtle interplay between
gravity, thermodynamics and quantum theory, which becomes apparent in
the presence of event horizons. The most famous instance of
this interplay is, of course, the Bekenstein-Hawking formula for
black hole temperature and entropy.

In brief, what we found is the following: If one describes the
process in terms of the so-called global coordinates for de Sitter
space, in which the volume of the spatial sections is proportional
to the hyperbolic cosine of time, the nucleation of membranes
occurs through the exact analog of the ($3+1$)-dimensional
instanton, as it was already noted in \cite{BT}. On the other
hand, if one describes the process in terms of the static
coordinates for de Sitter space, the membranes are nucleated
through the thermalon, that is a classical thermodynamical effect.
It is to be emphasized that when we say ``... describes in terms
of coordinates ...", we are talking about just a change of
coordinates in one and the same spacetime : two portions of de
Sitter spaces of different radii of curvature joined across the
membrane.

Thus, one and the same physical
process appears in one description as a strictly
quantum-mechanical tunneling effect  and, in the other, as a classical metastability effect 
(going over the barrier rather than tunneling
through it). There is no paradox here because one may say that
when the coordinates are changed, the ``potential" is also changed
so the barrier that one tunnels through in one case is not the
same barrier over whose top one goes  in the other. It is also
 quite alright that the thermodynamic description should appear when the
static coordinates are used, since it would appear natural to think that
time-independence amounts to equilibrium.

\section{Instantons and thermalons}

\begin{figure}[h]
\centering
\includegraphics[width=8cm]{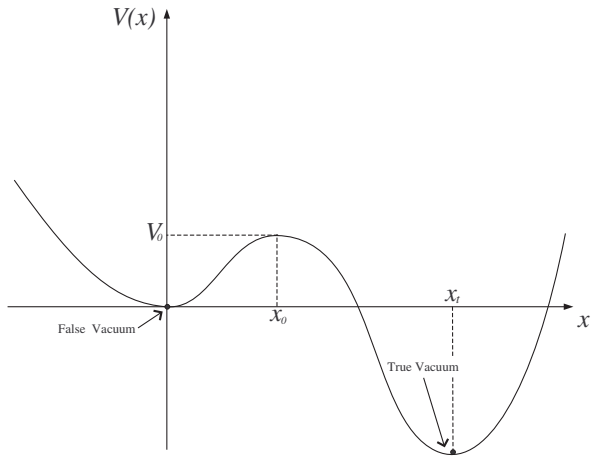}
\caption{A particle in a potential is initially in a metastable state at $x=0$ (false vacuum). It may end up in
the true vacuum at $x=x_t$ either by tunneling through the potential barrier or by a thermal kick.}
\label{lorentz}
\end{figure}

To make the discussion self-contained we briefly review here the
concept of ``thermalon". We also compare and
confront it with the concept of instanton. Consider a non-relativistic 
one-dimensional
particle placed in a potential $V(r)$, as shown in Fig.
\ref{lorentz}. The particle is initially in a metastable
minimum (the ``false vacuum"), at $x=0$. The particle may decay to
the ``true vacuum" at $x=x_t$ by two different mechanisms: It can
either (i) tunnel through the potential barrier or 
(ii) it can go over the barrier by a thermal kick.

\begin{figure}[h]
\centering
\includegraphics[width=8cm]{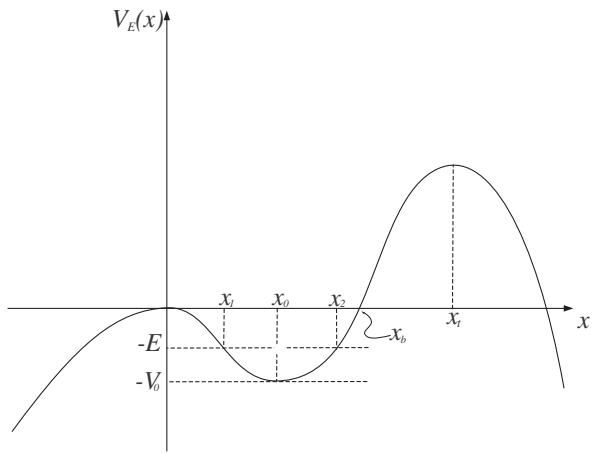}
\caption{The Euclidean system consists in the particle placed in a potential $V_E=-V$. 
The Euclidean path integral is evaluated in the semiclassical limit. At zero temperature,
the solution that dominates the path integral is the instanton which starts at rest at  $x=0$, $t=-\infty$, bounces at 
$x=x_b$ and comes back to $x=0$ at $t=\infty$. At higher 
inverse temperatures $\beta$, the corresponding instantons are periodic orbits, bounded between two turning
points, $x_1$, $x_2$, with period $\beta\hbar$. This happens at some
particular value of the energy $-E$. When the temperature is high enough the
static orbit at the bottom of the potential well $x=x_0$, the thermalon, becomes dominant.}
\label{instanton}
\end{figure}

For the tunneling problem, if one were to treat the problem exactly one would start with a wave packet
localized around $x=0$ and calculate  the quantum mechanical amplitude
to propagate from $x=0$ to $x=x_t$. This one could do
in principle at any arbitrary inverse temperature $\beta$ by computing the
path integral.
In the semiclassical approximation the path integral is dominated by
a solution of the classical equations of motion in imaginary time 
(``Wick rotation", ``Euclidean continuation"). The calculation is further
simplified at zero or very low temperatures $\beta^{-1}$.

For the thermal excitation the solution is again given by an Euclidean
path integral (the partition function), which now is most simply dealt 
with for $\beta^{-1}$ large enough to go over the top of the barrier by a
``thermal kick". For tunneling,
the classical solution is called the instanton and it is described in 
Fig. \ref{instanton}. The particle starts at rest from the false vacuum at $t=-\infty$, 
bounces at $x=x_b$ and comes back to the false vacuum at $t=\infty$.
The decay rate takes the form \cite{coleman,affleck}
\begin{equation}
\Gamma = \frac{2}{\hbar}\mbox{Im}F \ \ \ \ \ \ \ \ \
\mbox{(quantum
tunneling)}. 
\label{q}
\end{equation}
For a thermal kick the classical solution is called a thermalon. The particle sits at all times at the stable 
equilibrium position $x=x_0$ of the Euclidean potential of Fig. \ref{instanton}. 
The decay takes the form \cite{langer,affleck}
\begin{equation}
\Gamma = \frac{\omega\beta}{\pi}\mbox{Im}F  \ \ \ \ \ \
\mbox{(thermal activation)} .
\label{t}
\end{equation}
In both  Eq. (\ref{q}) and Eq. (\ref{t}), $F$ is the Helmholtz free energy given
by
\begin{equation}
\beta F = \beta U - S,
\label{helmholtz}
\end{equation}
where $U$ is the internal energy and $S$ the entropy.
In Eq. (\ref{t}), $\omega$ is the frequency of oscillations in Euclidean time around
$x=x_0$ (Fig.(\ref{instanton})),
\begin{equation}
\omega = V^{\prime\prime}_{E}(x_0) .
\label{o}
\end{equation}

If one evaluates the path integral 
to imaginary time,{\it i.e.}
\begin{equation}
\label{im}
x(t)=x_E(it_E), \ \ \ \ \ \ iI[x(t)] = I(x_E(t_E))  ,
\end{equation}
one has
\begin{equation}
\int {\cal D}x_E e^{\frac{i}{\hbar} I_{E}} = e^{-\beta F},
\label{path}
\end{equation}
where ``$E$" stands for ``Euclidean" and the functional integral is evaluated over closed paths
$x_E(t_E)$ with Euclidean time period
\begin{equation}
\Delta t_E= \hbar\beta .
\label{et}
\end{equation}

For tunneling as well as for thermal activation, the semiclassical rates take
the form,
\begin{equation}
\label{rate} \Gamma = Ae^{-B},
\end{equation}
where $B$ is the classical Euclidean action
divided by $\hbar$ and $A$ is a prefactor which involves
the determinant of a differential operator \cite{coleman}, and
which will not be discussed here.

The properties of instantons and thermalons are illustrated and
discussed in Fig.\ref{instanton} and summarized in Table I \cite{affleck,linde}.
\begin{table}
\label{table}
\caption{Instantons and Thermalons Compared and Contrasted.}
\begin{small}
\begin{tabular}{|p{2.5cm}||p{3.5cm}|p{3.5cm}|p{4cm}|p{4cm}|}\hline
Solution of the classical  Equation of motion & 
Process that the solution describes when used to 
dominate path integral& 
Dependence on Euclidean
time for motion in one-dimensional $\;\;\;\;$ potential&Range of validity if used to approximate
decay rate  by steepest descent&Formula for decay rate\\
\hline\hline
Instanton at zero
temperature &Tunneling through potential barrier&Time dependent.
Starts at $x_1$  at $t=-\infty$, bounces at $x_2$ and comes back to $x_1$ at $t=\infty$.
The bounce is localized in time. 
& Zero temperature & $$\Gamma=\frac{2}{\hbar}\mbox{Im}F=\frac{2}{\hbar}\mbox{Im}E$$ \\
Instanton at non-zero temperature & Tunneling through potential barrier
at non-zero temperature &Time dependent, starts at $x_1$, bounces at $x_2$ and comes back
to $x_1$ after a full period $\Delta t=\beta$. & Dominates for 
$$\frac{2\pi}{\omega}\ll\beta\hbar<\infty$$& $$\Gamma=\frac{2}{\hbar}\mbox{Im}F$$\\
Thermalon&Going over the barrier by a thermal kick& Time independent.$\; \; \;$ Stands at
the bottom of the Euclidean potential for all times.& Dominates for 
$$0<\beta\hbar\ll\frac{2\pi}{\omega}$$& $$\Gamma = \frac{\omega\beta}{\pi}\mbox{Im}F$$ \\
& & & In overlap region, $\beta\hbar\sim\frac{2\pi}{\omega}$, both solutions 
should in general be included. 

Here $\omega$ is the frequency of oscillations
in Euclidean time of perturbations around the thermalon. It is assumed that there is only one
such stable Euclidean mode. For motion in a potential, $\omega=V^{\prime\prime}(x_T)$.
& $F$ has imaginary part because the extremum is a saddle point due to unstable state.\\ \hline
\end{tabular}
\end{small}
\end{table}
\section{Pair creation in flat spacetime}

\subsection{Instanton}

In two spacetime dimensions, a closed membrane, which may be
thought of as the boundary of a ball, becomes a pair of points,
which are the boundary of an interval. Thus, we will be
considering pair creation. The term ``pair creation" is all the
more appropriate since the analog of the $3$-form potential
appearing in four spacetime dimensions is, in two spacetime
dimensions, just the ordinary electromagnetic potential.
Therefore, our problem is pair creation by an electric field
coupled to gravity. The action will be taken to be the sum of four terms:
(i) The length of the worldline times the mass of the particle, (ii) the
minimal coupling to the electromagnetic field, (iii) the Maxwell action,
(iv) the gravitational action, for which we will use the 
functional proposed in  \cite{CT}. However, for the sake of focusing 
as clearly as possible on the central point, we shall start by considering
pair creation in flat spacetime in a constant external electric field $E$. This is of
interest because even this simplified process  accepts 
the two alternative interpretations, namely, quantum mechanical or thermodynamical
depending on the coordinate system we choose in our description.

The Euclidean action, describing a particle of charge $q$ and mass $m$,
with worldline parameterized by $z^{\mu}(\lambda)$, in an external electromagnetic field
$A_\mu=(E x,0)$ is 
\begin{equation}
I=m\int\sqrt{\dot{z}^{\mu}\dot{z}_{\mu}}d\tau - q\int A \
= m(\mbox{length}) - qE(\mbox{area})  .
\label{action}
\end{equation}
The overall sign of the
action has been chosen so that one path integrates $e^{-I}$.

In Cartesian coordinates, the action (\ref{action}) reads
\begin{equation}
I=\int d\lambda \left[   m\sqrt{\dot{t}^2+\dot{x}^2} -qEx\dot{t}    \right] ,
\label{decartes}
\end{equation}
where $\lambda$ is a parameter that increases along the worldline. The momentum 
$P_t$ conjugate to $t$ is given by
\begin{equation}
P_t = m \frac{\dot{t}}{\sqrt{\dot{t}^2+\dot{x}^2}}  - qEx
=\mbox{sgn}(\dot{t})\frac{m}{\sqrt{1+\left(\frac{dx}{dt}\right)^2}} - qEx  .
\label{P0}
\end{equation}

\begin{figure}[h]
\centering
\includegraphics[width=8cm]{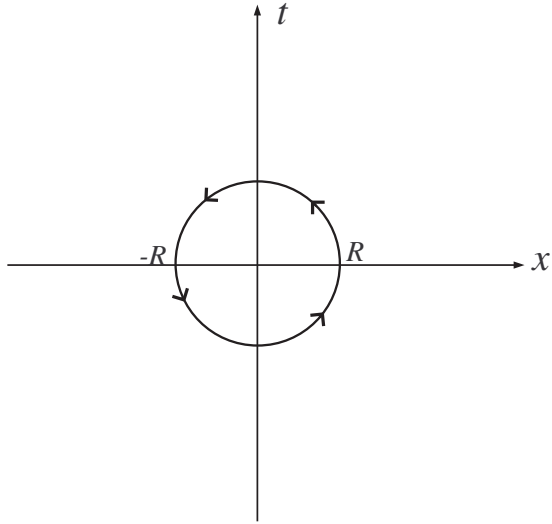}
\caption{The instanton is an oriented  
circle of radius $R=m/qE$ centered at any point $(t_0,x_0)$ of the Euclidean plane.}
\label{circle}
\end{figure}

The instanton solution is a complete circle in the $(x,t)$--plane centered at $(t_0,x_0)$
with radius $R$ equals to
\begin{equation}
R=m/qE,
\label{R}
\end{equation}
where $t_0$ and $x_0$ are integration constants (See Fig \ref{circle}).
One may describe it as follows. The system 
is initially in the metastable vacuum (no particle). At $t_0-R$, a particle-antiparticle pair appears. The particles 
then propagate and annihilate at $t_0+R$, leaving back the metastable vacuum.

Note that this is the most general solution
of the equations of motion.  This is an instanton at zero temperature, because
it is time dependent and localized in time (see Table I). The instanton remains a solution
if one identifies Euclidean time with a period $\hbar\beta$ provided the circle of Fig. \ref{circle}
fits into the corresponding cylinder, i.e.,
\begin{equation}
2R<\hbar\beta .
\label{cylinder}
\end{equation}
It is then an instanton at non-zero temperature $\beta^{-1}$.

The action evaluated on this orbit is
\begin{equation}
I_{0}=\frac{\pi m^2}{qE} \ ,
\label{I0}
\end{equation}
which is the Schwinger\cite{schwinger} result for two--dimensional spacetime.

The  tunneling decay rate described by the instanton is, in the
semiclassical approximation,
\begin{equation}
\Gamma \sim e^{-I_{0}/\hbar} = e^{-\frac{m^2\pi}{qE}}\ .
\label{dos}
\end{equation}

\subsection{Thermalon}

 The above interpretation changes radically if
we use  polar coordinates,
\begin{eqnarray}
t &=& r\sin(\tau/L) \\
x &=& r\cos(\tau/L) \ .
\end{eqnarray}
Here $L$ is some arbitrary length scale. In the Lorentzian continuation,
these are Rindler coordinates adapted to an observer of proper acceleration
$1/L$. The metric takes the form
\begin{equation}
ds^2 = \frac{r^2}{L^2}d\tau^2 + dr^2 \ ,
\label{rindlee}
\end{equation}
The acceleration of a stationary particle at $r$ is $1/r$. The observer
located at $r=L$ has the special property that on his trajectory  
the Killing vector $d/d\tau$ has unit norm.

The Euclidean Rindler ``time"  coordinate $\tau$  has a periodicity $2\pi L$ and
is identified with the inverse Unruh temperature seen by the associated
Lorentzian accelerated observer at $r=L$,
\begin{equation}
\beta =\frac{2\pi L}{\hbar} \ .
\label{temp}
\end{equation}

The action (\ref{action}) now reads,
\begin{equation}
I = \int d\lambda\left[ m \sqrt{\dot{r}^2+\frac{r^2}{L^2}\dot{\tau}^2}
- \frac{qE}{2L} r^2 \dot{\tau} \right]  .
\label{actionpolar}
\end{equation}
We evaluate the conserved momentum $P_{\tau}$ conjugate to $\tau$,
\begin{equation}
P_{\tau}= \frac{m\dot{\tau}}{\sqrt{\frac{L^2}{r^2}\dot{r}^2+1}}- \frac{Eqr^2}{2L}
=\frac{m \ \mbox{sgn}(\dot{\tau})}{\sqrt{\frac{L^2}{r^2}\left(\frac{dr}{d\tau}\right)^2+1}}- \frac{Eqr^2}{2L}.
\label{Ptau}
\end{equation}

Contrary to what happens in Cartesian coordinates, the dynamical system admits now static solutions.  
In fact, there is only one such solution, given by  
\begin{equation}
r=R=\frac{m}{E q},
\label{r}
\end{equation}
and
\begin{equation}
P_\tau= \frac{1}{2} \frac{m^2}{LEq}= \frac{1}{2}m\left(\frac{R}{L}\right)  .
\label{mom}
\end{equation}
Because it is static and stable (see below), this solution is a thermalon.
The value of $R$ given by (\ref{r}) is, of course, just Eq. (\ref{R}). 
This was expected, and it is the main point that we are making, namely, that 
the instanton as seen in a
polar system of coordinates, centered a the origin $t_0=x_0=0$, is the thermalon.

A small perturbation of the solution amounts to translating the center of the circle.  
When viewed in Rindler coordinates, this appears as a periodic motion with same period 
as $\tau$.  So, the solution is stable and 
the frequency of oscillations around the thermalon is,
\begin{equation}
\omega_{\tau}= \frac{1}{L} \ .
\label{freq}
\end{equation}

The thermalon
describes the probability for the particle to jump over the potential barrier by a thermal fluctuation.
The thermal decay rate is given by Eq. (\ref{rate}), which, in the thermalon case, reduces to the Boltzmann factor,
\begin{equation}
\Gamma \sim e^{-\beta {\cal E}} \ ,
\label{uno}
\end{equation}
where  the energy ${\cal E}$ is the conserved momentum $P_{\tau}$,
and $\beta$ the corresponding inverse Unruh temperature. Explicitly, one gets,
\begin{equation}
\beta {\cal E} = \frac{2\pi L}{\hbar}  \frac{1}{2} \frac{m^2}{LEq} = \frac{\pi m^2}{qE\hbar} .
\label{be}
\end{equation}
Note that although $\beta$ and ${\cal E}$ depend on $L$, this length scale drops out
from the product $\beta {\cal E}$. The value of $\beta {\cal E}$ given by Eq. (\ref{be})
is exactly the value of the action (\ref{I0}) for the instanton.
What underlies these ``coincidences" is, of course, that the thermalon is just the 
instanton described in a different coordinate system. However, we are not facing a 
triviality, because the concepts associated to each description are drastically different. 
In the instanton framework, we are analyzing a quantum mechanical 
process at zero temperature, while in the thermalon description, we are working out a 
classical thermodynamical instability at non-zero temperature.
Thus, we have in a very simple context, an example of the inextricable connection between 
gravity, quantum mechanics and thermodynamics, 
which was first observed  in the context of black hole entropy.

The equivalence of the two representations does not just happen
for the exponent in the transition rate, but in fact for the complete decay rates.
Indeed, if we multiply the temperature
(\ref{temp}) and the frequency (\ref{o}) and divide the product by
$\pi$ to evaluate the multiplicative factor characteristic of the
metastability calculation, Eq. (\ref{t}), we find
\begin{equation}
\frac{\omega_{\tau}\beta}{\pi}=\frac{1}{L}\frac{2\pi L}{\hbar}\frac{1}{\pi}=\frac{2}{\hbar}\ ,
\label{c}
\end{equation}
which is exactly the corresponding tunneling expression, Eq. (\ref{q}).

Eq. (\ref{c}) tells us that we are precisely in the overlap region where
both the thermalon and the instanton should be considered when computing
the free energy $F$ (see table I). However, in this example
there is only one Euclidean solution. It is just that it is 
interpreted differently in different coordinate systems. This unique solution
is the one that dominates the path integral in the present case.

As a final comment, we recall that in $D$ spacetime dimensions,
the symmetry group of the instanton at zero temperature is $SO(D)$, while the symmetry
group of the thermalon is $SO(D-1)\times SO(2)$. For the special case of
two dimensions the two groups coincide, in agreement with the fact that the
instanton and the thermalon are one and the same solution.

\section{Coupling to gravity}

The complementary description of one and the same physical process (pair creation)
as a quantum mechanical effect or as a thermal effect is also available when we switch
on gravity.
The actual calculations describing pair creation in a gravitational field
have already been done in \cite{BT}. As it was stated in the
introduction, the main novelty of the present work is the
complementary interpretation of the process as a thermal effect.

We will first remind the reader of the results of \cite{BT} in
the two--dimensional case and introduce
some  notation.
The equation of motion to be used for the gravitational field is \cite{CT}
\begin{equation}
\frac{1}{2}R-\lambda=\kappa T  \ \ ,
\label{1dimgr}
\end{equation}
where $T$ is the trace of the energy-momentum tensor, and
$\kappa$ a positive coupling (which would equal $8\pi G$ in four dimensions).
In our case, the energy momentum is the one produced by an
electromagnetic field $F_{\mu\mu}$ and   a particle
of mass $m$ and charge $q$. The Maxwell equations for the electromagnetic field
imply that $F_{\mu\nu}=\sqrt{g}E\epsilon_{\mu\nu}$, where $\epsilon_{\mu\nu}$
is the Levi-Civita symbol in two dimensions and $E$ is a constant in the absence of sources. 
The value of $E$ jumps when crossing the worldline of the particle, so that,
\begin{equation}
E_{+}-E_{-}=q \ .
\label{Maxwell}
\end{equation}
We will adopt the convention that, when traveling along the worldline
of the particle, the ``interior" ("-" region) will be on the right hand side.
The "+" region will be called the "exterior".  For the gravitational field,
the electromagnetic field will contribute to an effective
cosmological constant,
\begin{equation}
\Lambda_{\pm}\equiv \frac{1}{l_{\pm}^2} = \lambda + \kappa E_{\pm}^2   \ .
\label{cc}
\end{equation}
Therefore, at each side of the worldline, the geometry will be that of
a two--sphere of radii $l_+$ and $l_-$ respectively,
\begin{equation}
ds^2_{\pm}=l^2_{\pm}(d\theta^2+ \sin^2\theta_{\pm} d\phi_{\pm}^2)
\ \ \ \ (0\le\theta_\pm\le\pi, \ 0\le\phi_\pm\le2\pi) \ .
\end{equation}
By taking the range of $\phi$ to go from $0$ to $2\pi$, we are excluding
conical singularities at the poles. Note that we may always rotate the coordinate 
systems on each of the glued spheres so that when embedded in flat three-dimensional space 
their corresponding $z$--axis coincide. This means that
we can set $\phi_+=\phi_-\equiv\phi$. 
The worldline of the particle is determined by the matching
conditions obtained by integrating (\ref{1dimgr}) across the membrane. One gets
\begin{equation}
K_+ - K_- = \kappa m  \ ,
\label{israel}
\end{equation}
where $K_{\pm}$ are the extrinsic curvatures of the worldline
as embedded in each sphere (see for example \cite{BT}).

We now define 
\begin{equation}
\rho_{\pm}=l_\pm \sin\theta_\pm .
\label{rho}
\end{equation}
and parameterize the curve using the arclength $s$,
so that the extrinsic curvature is given by
\begin{equation}
K_{\pm}=\frac{1}{\rho_{\pm}\dot{\rho}_{\pm}}
\left(1-\frac{\rho^2_{\pm}}{l^2_{\pm}}\right)^{1/2}
\frac{d}{d\tau}\left(\rho_{\pm}^2\dot{\phi}_{\pm}\right) \ .
\label{ec}
\end{equation}

\begin{figure}[h]
\centering
\includegraphics[width=8cm]{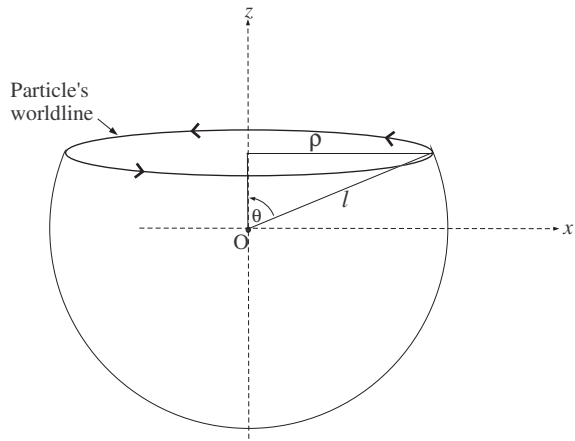}
\caption{Each de-Sitter two-sphere is parameterized as shown in the figure. 
The angle $\phi$ runs anti--clockwise around $z$. Instead of $\theta$ we may 
also use the coordinates $\rho=l\sin\theta$.  In this case, however, 
the parameterization will cover only half of the sphere.
A solution for the worldline of the particle has been drawn, and the ``exterior'' 
region has been removed.}
\label{four}
\end{figure}
 The trajectory is a circle which we take at some constant 
$\rho_\pm$ (see Fig. \ref{four}).
From Eqs. (\ref{israel}) and (\ref{ec}) we get,
\begin{equation}\label{israelrho}
\epsilon_+\sqrt{\frac{1}{\rho_+^2}-\frac{1}{l_+^2}} -
\epsilon_-\sqrt{\frac{1}{\rho_-^2}-\frac{1}{l_-^2}}= km ,
\end{equation}
where $\epsilon_\pm=\mbox{sgn}(\dot{\phi}_\pm)$.
Because the length of the circle must be the same as seen from each side, we must take
$\rho_-=\rho_+\equiv \bar{\rho}$ on the orbit. Eq. (\ref{israelrho}) yields then
 \begin{equation}
\bar{\rho}^2=\frac{l_+^2}{1+\gamma^2} \ ,
\label{rad}
\end{equation}
where
\begin{equation}
\alpha^2= \frac{1}{l_+^2}-\frac{1}{l_-^2} = 2\kappa q E_{av}\ ,
\label{alpha}
\end{equation}
\begin{equation}
E_{av}=\frac{1}{2}\left(E_+ + E_-\right) \ , 
\label{eav}
\end{equation}
and
\begin{equation}
\gamma=\frac{l_+(\alpha^2-\kappa^2m^2)}{2m\kappa} .
\label{gamma}
\end{equation}

Therefore, the classical Euclidean solution, the exponential of
whose action appears in the probability, consists of two
two-dimensional spheres of radii $l_+$ and $l_-$ joined at a
circle of radius $\bar{\rho}$.

\subsection{Instanton}

\begin{figure}[h]
\centering
\includegraphics[width=8cm]{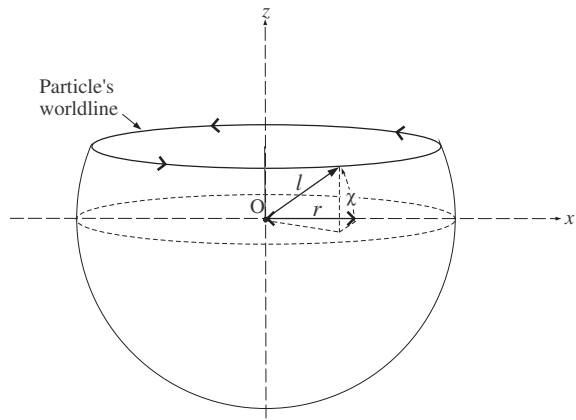}
\caption{Instanton interpretation. In each region we choose spherical coordinates such the
azimuthal angle $\chi$ is measured from the $x$ axis.  The angle $\xi_\pm$ goes 
anti-clockwise around the $x$ axis. In the radial coordinate $r=l\cos\chi$ the
metric takes the Schwarzschild-de Sitter form of Eq. (\ref{metricr}). The trajectory
of the particle $r(t)$ is time-dependent in the Euclidean time $t=l\xi$.} 
\label{five}
\end{figure}

We can choose different coordinate systems to 
describe the above solution. The interpretation of the solution
will depend on this choice. The instanton picture is obtained
by choosing spherical coordinates as in Fig. \ref{five}, i.e., 
\begin{equation}
ds^2_{\pm}=l^2_{\pm}(d\chi_\pm^2+ \sin^2\chi_{\pm} d\xi_{\pm}^2)
\ \ \ \ (0\le\chi_\pm\le\pi, \ 0\le\xi_\pm\le 2\pi)  .
\label{metricterm}
\end{equation}
The angle $\chi_\pm$ is now measured from
the $x$ axis, and $\xi_\pm$ goes anti-clockwise around the $x$ axis.
The Euclidean time is $t_\pm=l_\pm \xi_{\pm}$. The worldline of the particle is time-dependent and
turns out to be exactly the instanton solution discussed in \cite{BT}. 

To describe it more explicitly, define $r_\pm=l_\pm\cos\chi_{\pm}$. 
Again, the coordinates $r_+$ and $r_-$ must be related on
the worldline of the particle so that the total length is the same as seen from each side.
The relation may be obtained by noting that in terms of the previous coordinate system,
\begin{equation}
r_{\pm} = \rho_{\pm}\sin\phi ,
\label{}
\end{equation} 
Now, we know that $\rho_+=\rho_-$ on the trajectory, and therefore, we have
that $r_+=r_-\equiv r$. The metric takes the Schwarzschild-de Sitter form
\begin{equation}
ds^2_{\pm} = f^2_{\pm}dt^2_{\pm} + f_{\pm}^{-2}dr^2  ,
\label{metricr}
\end{equation}
with
\begin{equation}
f^2_\pm=1-\frac{r^2}{l_{\pm}^2} ,
\label{f}
\end{equation}
and the trajectory, as seen from each side, is given by
\begin{equation}
r(t)=\pm l\sqrt{\left( 1- \frac{\cos^2 \theta_0}{\cos^2{(t/l)}}  \right)}, \ \ \ \ \ \ 
\theta_0 \le t/l \le \theta_0 ,
\label{trajectory}
\end{equation}
where, for the sake of clarity, the subscripts $\pm$ have been dropped. The angle 
$\theta_0$ is a constant, which, for
each region, is given by Eqs. (\ref{rho}) and (\ref{rad}). 
As for the instanton in flat space, there are two points 
on the trajectory for each value of $t$.
Furthermore, if we take the limit 
$\kappa\rightarrow 0$
we end up with the instanton in flat spacetime of Sec. IIIA. 
To see this one has to recall that
the bare cosmological constant, $\lambda$, coming from ``the rest of physics"
in Eq. (\ref{1dimgr})  is of the form
\begin{equation}
\lambda=\kappa\rho_{vac}  ,
\label{ccenergy}
\end{equation}
where $\rho_{vac}$ is the energy density of the vacuum. When we 
take the limit, the cosmological constant $1/l^2_{\pm}$ goes to zero in both regions, 
\begin{equation}
\frac{1}{l_{\pm}}= \kappa\left(\rho_{vac} + E_{\pm}^2 \right) ,
\end{equation}
and therefore the metrics in Eq. (\ref{metricr}) become flat
Minkowski line elements in polar coordinates.
Furthermore, the radius of the orbit, given by Eq. (\ref{rad}),
goes into
\begin{equation}
\bar{\rho}^2=\frac{l_+^2}{1+\gamma^2} \stackrel{\kappa\rightarrow 0}{\longrightarrow}
\frac{1}{\frac{1}{l_+^2}+\frac{(E_+^2-E_-^2)^2}{4m^2}}
 \stackrel{\kappa\rightarrow 0}{\longrightarrow}
 \frac{m^2}{q^2E_{av}^2} \ ,
\label{}
\end{equation}
This is exactly the radius we obtained in the flat 
space case in Eq.(\ref{R}).

\subsection{Thermalon}

\begin{figure}[h]
\centering
\includegraphics[width=5cm]{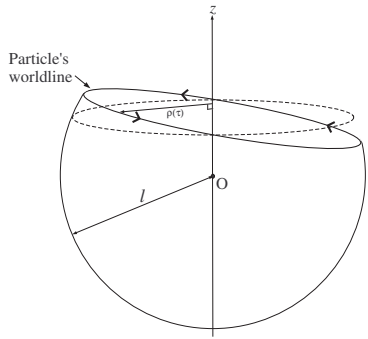}
\caption{The diagram depicts a small perturbation on the thermalon alignment of Fig. \ref{four}. The radius $\rho(\tau)$ oscillates
with a period equal to $2\pi l$, the total period of the time coordinate $t$. }
\label{two}
\end{figure}

The thermalon picture is obtained by analyzing the trajectory in the original  coordinate 
system of Fig. \ref{four}. The solution is then clearly static. 
The metric takes the form
\begin{equation}
ds^2_{\pm}=\frac{d\rho^2}{   1-\frac{\rho^2}{l^2_{\pm}}    }
+ \frac{\rho^2}{l^2_{\pm}} dt_{\pm}^2  \ ,
\label{mrho}
\end{equation}
with the ``Euclidean time" $t_\pm=l_\pm\phi$. 
It is interest to see at this point how one may recover the flat 
space background analysis of the previous section as a limiting case
of the present discussion. Taking the limit $\kappa\rightarrow 0$
we must define a new, rescaled time coordinate $\tau$,
\begin{equation}
t_{\pm} = \frac{l_{\pm}}{L}\tau  ,
\label{}
\end{equation}
with $L$ being an arbitrary constant with dimension of length.
In the limit, we get flat spacetime in polar coordinates
as in Eq. (\ref{rindlee}).

In the coordinate system used in Fig. \ref{four},
the particle remains at a fixed value of $\rho$ at all times. However,
if we  rotate the worldline of the particle as in Fig. \ref{five}, the new solution will not be static in
this set of coordinates. The curves ($\rho_+(\tau),t_+(\tau)$) and
($\rho_-(\tau),t_-(\tau)$) as seen from each side of the particle's trajectory  will oscillate
around the static solution with frequency 
\begin{equation}
\omega_{\pm}=1/l_{\pm}, 
\end{equation}
since, in the absence of a conical singularity, the period of the time coordinate in (\ref{mrho})
must be $2\pi l_\pm$.
Therefore, the solution  has the two key properties
of a thermalon: Is a static and stable Euclidean solution of the theory.

We now may identify the temperature felt by an observer at 
each region with  the respective de--Sitter temperature,
\begin{equation}
T_{\pm} = \frac{1}{\beta_{\pm}} =\frac{\hbar}{2\pi l_{\pm}}
\end{equation}
and immediately note that,
\begin{equation}
\frac{\omega_{\pm}\beta_{\pm}}{\pi}\frac{1}{l_{\pm}}\frac{2\pi l_{\pm}}{\hbar}\frac{1}{\pi}=\frac{2}{\hbar}\ .
\label{}
\end{equation}
We see that, again, the two expressions for the probability (\ref{q}) and (\ref{t})
coincide.

To obtain an expression for the decay rate, we need to compute
Im$F$ in expression (\ref{q}) or (\ref{t}), which, in our case are
equivalent. This has been done before for the instanton, and, as have shown
in the present work, both the instanton and the thermalon are one and the same
process in two dimensions. For the sake of completeness we will quote here the
known expressions.
The result is of the form (\ref{rate}). The value of the action $B$ was
computed in \cite{BT} and takes the value,
\begin{equation}
I=4\pi\bar{\rho}m-\frac{2\pi}{\kappa}\log\left(1+ 
  \frac{\kappa m l_+}{(\sqrt{1+\gamma^2}-\gamma)(1-\alpha^2l_+^2)} - \alpha^2 l_+^2 \right),
\label{B}
\end{equation}
with $\alpha^2$, $\gamma$ defined in (\ref{alpha}) and (\ref{gamma}) respectively.

\acknowledgments

This work  was funded by an institutional grant to CECS of the
Millennium Science Initiative, Chile, and Fundaci\'on Andes,
and also benefits from the
generous support to CECS by Empresas CMPC. AG gratefully
acknowledges support from FONDECYT grant 1010449  and from
Fundaci\'on Andes. AG and CT  acknowledge partial support under
FONDECYT grants 1010446 and 7010446.  
The work of MH is partially supported by
the ``Actions de Recherche Concert{\'e}es" of the ``Direction de
la Recherche Scientifique - Communaut{\'e} Fran{\c c}aise de
Belgique", by IISN - Belgium (convention 4.4505.86), by a ``P\^ole
d'Attraction Universitaire" and by the European Commission
programme MRTN-CT-2004-005104, in which he is associated to 
V.U. Brussel.

\end{document}